\documentstyle[preprint,aps]{revtex}  

\newcommand{\beq}{\begin{equation}}
\newcommand{\eeq}{\end{equation}}
\newcommand{\snu}{\tilde \nu}

\newcommand{\msnuone}{m_{\snu_+}}
\newcommand{\msnutwo}{m_{\snu_-}}

\newcommand{\dmsnu}{{\mbox{$\Delta m_{\tilde \nu}$}}}
\newcommand{\dmsnutwo}{{\mbox{$\Delta m^2_{\snu}$}}}

\newenvironment{Eqnarray}%
         {\arraycolsep 0.14em\begin{eqnarray}}{\end{eqnarray}}
\def\beqa{\begin{Eqnarray}}
\def\eeqa{\end{Eqnarray}}

\def\hh{H^0}
\def\hl{h^0}
\def\ha{A^0}
\def\mhh{m_{\hh}}
\def\mhl{m_{\hl}}
\def\mha{m_{\ha}}

\def\msnusnumn{(M^2_{\snu\snu^*})_{mn}}

\def\l{\lambda}

\def\lp{\lambda^\prime}

\def\ifmath#1{\relax\ifmmode #1\else $#1$\fi}
\def\eighth{\ifmath{{\textstyle{1 \over 8}}}}

\def\quarter{\ifmath{{\textstyle{1 \over 4}}}}

\def\Ref#1{ref.~\cite{#1}}

\def\eq#1{eq.~(\ref{#1})}

\def\eqs#1#2{eqs.~(\ref{#1}) and (\ref{#2})}

\def\npb#1{{\sl Nucl.\ Phys.}\ {\bf B#1}}
\def\plb#1{{\sl Phys.\ Lett.}\ {\bf B#1}}
\def\prd#1{{\sl Phys.\ Rev.}\ {\bf D#1}}
\def\prl#1{{\sl Phys.\ Rev.\ Lett.} {\bf #1}}

\def\epjc#1{{\sl Eur.~Phys.~J.}\ {\bf C#1}}

\def\ifmath#1{\relax\ifmmode #1\else $#1$\fi}
\def\half{\ifmath{{\textstyle{1 \over 2}}}}

\def\eighth{\ifmath{{\textstyle{1 \over 8}}}}
\def\cw{c_W}
\def\sw{s_W}

\begin{document}        
\baselineskip 14pt

\preprint{
\vbox{
      \hbox{SLAC-PUB-8173}
      \hbox{SCIPP-99/24}
      \hbox{hep-ph/9906310}
      \hbox{June, 1999}
}}

\bigskip
\bigskip
\title{Neutrino masses and sneutrino mixing 
in R-parity violating supersymmetry%
{\tighten%
\footnote{Invited talk presented by Yuval Grossman at the
American Physical Society (APS) meeting of the Division of Particles 
and Fields (DPF99).}%
\footnote{YG is supported by the U.S. Department of Energy under
contract DE-AC03-76SF00515, and
HEH is supported in part by the U.S. Department of Energy
under contract DE-FG03-92ER40689.}}
}
\author{Yuval Grossman\,$^a$ and  Howard E. Haber\,$^b$}
\address{
$^a$Stanford Linear Accelerator Center, 
        Stanford University, Stanford, CA 94309 \\
  $^b$Santa Cruz Institute for Particle Physics, 
University of California, Santa Cruz, CA 95064}


\maketitle              
{\tighten
\begin{abstract}        
R-parity-violating supersymmetry with a conserved baryon number $B$
provides a framework for particle physics with lepton number ($L$)
violating interactions.  Two important probes of the $L$-violating
physics are neutrino masses and sneutrino-antisneutrino
mass-splittings.  We evaluate these quantities in the context of the
most general CP-conserving, R-parity-violating $B$-conserving
extension of the minimal supersymmetric standard model.  In generic
three-generation models, three sneutrino-antisneutrino mass splittings
are generated at tree-level.  In contrast, only one neutrino mass is
generated at tree-level; the other two neutrinos acquire masses at
one-loop.  In many models, the dominant contribution to the radiative
neutrino masses is induced by the non-zero sneutrino-antisneutrino
mass splitting.
\end{abstract}   	
}
\newpage

\section{Introduction}
The solar and atmospheric neutrino anomalies provide strong indications
that the neutrinos are massive. In particular, the data
suggest that there is near-maximal mixing between $\nu_\mu$ and
$\nu_\tau$ but that their masses are hierarchically separated \cite{pdg}.
To accommodate this data, the Standard
Model must be extended, either by introducing right-handed neutrinos or
by adding Majorana neutrino mass terms that violate lepton number by two
units.

In low-energy supersymmetric extensions of the Standard Model, lepton
number and baryon number 
conservation is not automatically respected by the most general
set of renormalizable interactions.  However, the constraints on
baryon number violation are extremely severe in order to avoid fast
proton decay.  If one wants to enforce lepton number and baryon number
conservation in the tree-level supersymmetric theory, it is sufficient
to impose one extra discrete symmetry.  In the minimal
supersymmetric standard model (MSSM), a multiplicative symmetry called
R-parity is introduced, where the R quantum number of
an MSSM field of spin $S$, baryon number $B$ and lepton number $L$ 
is given by $(-1)^{[3(B-L)+2S]}$. By introducing this symmetry,
one eliminates all dimension-four lepton number 
and baryon number-violating interactions.  Majorana neutrino masses can
be generated in an R-parity-conserving extension of the MSSM 
involving new $\Delta L=2$ interactions through
the supersymmetric see-saw mechanism \cite{susyseesaw,GH}.  

Such $\Delta L=2$ interaction have an important impact
on sneutrino phenomena \cite{GH,GHRPV,HKK}.
The sneutrino ($\snu$) and antisneutrino ($\bar{\snu}$),
which are eigenstates of lepton number, are no longer mass eigenstates.  
The mass eigenstates are therefore superpositions of $\snu$ and $\bar{\snu}$,
and sneutrino mixing effects can lead to a phenomenology analogous to
that of $K$--$\overline K$ and $B$--$\overline B$ mixing. 
The mass splitting between the two sneutrino mass eigenstates is
related to the
magnitude of lepton number violation, which is typically characterized
by the size of neutrino masses. (In some cases the sneutrino mass 
splitting may be enhanced by a factor as large as $10^3$ compared 
to the corresponding neutrino mass \cite{GH,HMM}.)
As a result, the sneutrino mass splitting is expected generally to be
very small. Yet, it can be detected
in many cases, if one is able to observe the lepton number 
oscillation \cite{GH}.

The primary motivation for introducing a conserved R-parity
above was to impose a conserved baryon number to avoid fast proton
decay.  However, this can also be achieved in
low-energy supersymmetric models where $B$ is conserved
but $L$ is violated (so that R-parity is also violated).
In this paper, we focus on the $B$-conserving
R-parity-violating (RPV) extension of the MSSM.  In such a model,
neutrinos are massive \cite{all,bgnn,enrico,Hem,Ferr}
and sneutrino--antisneutrino pairs are no longer
mass-degenerate \cite{GH,GHRPV,HKK}.  
In Section II, we introduce the most general RPV extension of the MSSM 
with a conserved baryon number and establish our notation. 
In Section III, we show how a
tree-level mass for one neutrino is generated due to neutrino--neutralino 
mixing.
In Section IV, we exhibit how tree-level mass splittings for the three
sneutrino--antisneutrino pairs are generated due to sneutrino--Higgs bosons 
mixing.
In Section V, we calculate the neutrino masses 
generated at one loop. In Section VI, we argue that in many models,
the dominant contribution to the one-loop
neutrino masses is induced by the non-zero
sneutrino--antisneutrino mass splittings.
A brief summary and conclusions are given in Section VII.

\section{R-parity violation formalism}
In RPV low-energy supersymmetry, there is no conserved
quantum number that distinguishes the lepton supermultiplets $\hat
L_m$ and the down-type Higgs supermultiplet $\hat H_D$.  Here,
$m$ is a generation label that runs from 1 to $3$.  Each
supermultiplet transforms as a $Y=-1$ weak doublet under the electroweak
gauge group.  It is therefore convenient to denote the four
supermultiplets by one symbol $\hat L_\alpha$ ($\alpha=0,1,2,3$).
The Lagrangian of the theory is fixed by the superpotential and the
soft-supersymmetry-breaking terms. 

The most general renormalizable superpotential
respecting baryon number is given by:
\beq \label{rpvsuppot}
W=\epsilon_{ij} \left[
-\mu_\alpha \hat L_\alpha^i \hat H_U^j + 
\half\l_{\alpha\beta m}\hat L_\alpha^i \hat L_\beta^j \hat E_m +
\lp_{\alpha nm} \hat L_\alpha^i \hat Q_n^j  \hat D_m
-h_{nm}\hat H_U^i \hat Q^j_n \hat U_m
\right]\,,
\eeq
where $\hat H_U$ is the up-type Higgs supermultiplet, the
$\hat Q_n$ are doublet quark supermultiplets, $\hat U_m$ [$\hat D_m$] are 
singlet up-type [down-type] quark supermultiplets
and the $\hat E_m$ are the singlet charged lepton 
supermultiplets.
(Our notational conventions follow those of \Ref{GHRPV}.)
Without loss of generality, the coefficients $\lambda_{\alpha\beta m}$ 
are taken to be antisymmetric under the interchange of the indices
$\alpha$ and $\beta$. Note that   
the $\mu$-term of the MSSM is now extended to an 
$4$-component vector, $\mu_\alpha$.
Then, the trilinear terms in the superpotential proportional to
$\lambda$ and $\lambda'$
contain lepton number violating generalizations of the down quark
and charged lepton Yukawa matrices.

Next, we consider the most general set of (renormalizable) 
soft-supersymmetry-breaking terms.  In addition to the usual 
soft-supersymmetry-breaking terms of the R-parity-conserving MSSM, one
must also add new $A$ and $B$ terms corresponding to the RPV
terms of the superpotential.  In addition, new RPV scalar squared-mass
terms also exist.  As above, we can streamline the notation by
extending the definitions of the coefficients of the R-parity-conserving
soft-supersymmetry-breaking terms to allow for an index of type
$\alpha$ which can run from 0 to 3 (while Latin indices $m$~and~$n$ run from 
1 to 3). Explicitly,
\beqa \label{softsusy}
 V_{\rm soft}  &=&  (M^2_{\widetilde Q})_{mn}\,\widetilde Q^{i*}_m
        \widetilde Q^i_n 
   +  (M^2_{\widetilde U})_{mn}\,\widetilde U_m^*\widetilde U_n        
   +  (M^2_{\widetilde D})_{mn}\,\widetilde D_m^*\widetilde D_n
            \nonumber \\
 && +  (M^2_{\widetilde L})_{\alpha\beta}\,
          \widetilde L^{i*}_\alpha\widetilde L^i_\beta
      + (M^2_{\widetilde E})_{mn}\,\widetilde E_m^*\widetilde E_n
       + m^2_U|H_U|^2      
  -(\epsilon_{ij} b_\alpha\tilde L_\alpha^i H_U^j +{\rm h.c.}) \nonumber\\
 && +  \epsilon_{ij} \bigl[\half a_{\alpha\beta m} \widetilde L^i_\alpha
       \widetilde L^j_\beta \widetilde E_m + a'_{\alpha nm}
       \widetilde L^i_\alpha\widetilde Q^j_n\widetilde D_m
       - (a_U)_{nm} H^i_U
        \widetilde Q^j_n\widetilde U_m + {\rm h.c.}\bigr]\nonumber \\
 && +  \half \left[ M_3\, \widetilde g
   \,\widetilde g + M_2 \widetilde W^a\widetilde W^a
  + M_1 \widetilde B \widetilde B +{\rm h.c.}\right]\,.
\eeqa
Note that the
single $B$ term of the MSSM
is extended to a $4$-component vector $b_\alpha$, and
the squared-mass term for the down-type Higgs boson
and the $3\times 3$ lepton scalar squared-mass 
matrix are combined into a $4 \times 4$ matrix. 
In addition, the matrix $A$-parameters
of the MSSM are extended in the obvious manner [analogous to the
Yukawa coupling matrices in \eq{rpvsuppot}]; in particular, 
$a_{\alpha\beta m}$ is antisymmetric under the interchange of
$\alpha$ and $\beta$.  
It is sometimes 
convenient to follow the more conventional notation in the literature
and define the $A$ and $B$ parameters as follows:
\beqa \label{abterms}
&&a_{\alpha\beta m}\equiv \lambda_{\alpha\beta m} (A_E)_{\alpha\beta m}
\,,\qquad
(a_U)_{nm}\equiv h_{nm} (A_U)_{nm}\,,\nonumber \\
&&a'_{\alpha nm}\equiv \lambda'_{\alpha nm} (A_D)_{\alpha nm}\,,\qquad
b_\alpha\equiv \mu_\alpha B_\alpha\,,
\eeqa
where repeated indices are not summed over in the above equations.
Finally, the Majorana gaugino
masses, $M_i$, are unchanged from the MSSM.

The total scalar potential is given by:
\beq \label{scalarpot}
V_{\rm scalar}=V_F+V_D+V_{\rm soft}\,,
\eeq
where explicit forms for the supersymmetric contributions 
$V_F$ and $V_D$ can be found in \Ref{GHRPV}.
We do not present here the minimization conditions, but only mention that 
regions of parameter space exist where only the neutral color-singlet
scalar fields acquire vacuum expectation values: $\langle
h_U\rangle\equiv v_u/\sqrt{2}$ and $\langle\snu_\alpha\rangle\equiv 
v_\alpha/\sqrt{2}$.

Up to this point, there is no preferred direction in the generalized
generation space spanned by the $\hat L_\alpha$.  However,
It is sometime convenient to choose a particular ``interaction'' basis
such that $v_m=0$, in which case $v_0=v_d$.
In this basis, we denote $\hat L_0\equiv\hat H_D$.  

For simplicity, we shall impose CP-invariance on the model (which
implies that all parameters appearing in \eqs{rpvsuppot}{softsusy}
and the vacuum expectation values of the scalar fields can be taken to
be real).  The consequences of CP-violating effects in this model will
be considered elsewhere.

\section{Neutrino mass at tree level}
The neutrino can become massive due to mixing with the neutralinos \cite{all}.
This is determined by the $7 \times 7$ mass matrix 
in a basis spanned by the two
neutral gauginos $\widetilde B$ and $\widetilde W^3$, the higgsinos
$\widetilde h_U$ and $\widetilde h_D\equiv\nu_0$,
and 3 generations of neutrinos, $\nu_m$.  The tree-level 
fermion mass matrix, with rows and columns corresponding to
$\{\widetilde B,\widetilde W^3,\widetilde h_U, 
\nu_\beta\}$ is given by \cite{bgnn,enrico}:
\beq
M^{\rm (n)}=\pmatrix{
M_1&0&m_Z\sw v_u/v&-m_Z\sw v_\beta/v\cr
0&M_2&-m_Z\cw v_u/v&m_Z\cw v_\beta/v\cr
m_Z\sw v_u/v&-m_Z\cw v_u/v&0&\mu_\beta\cr
-m_Z\sw v_\alpha/v&m_Z\cw v_\alpha/v&\mu_\alpha&0_{\alpha\beta}\cr}\,,
\eeq
where $\cw\equiv\cos\theta_W$, $\sw\equiv\sin\theta_W$,
and $0_{\alpha\beta}$ is the $4\times 4$
zero matrix.  In a basis-independent analysis, 
it is convenient to introduce:
\beq\label{xidef}
\cos\xi\equiv {\sum_\alpha v_\alpha\mu_\alpha\over v_d\mu},
\eeq
where 
\beq \label{mudef}
\mu^2\equiv \sum_\alpha\mu_\alpha^2\,, \qquad
v_d^2\equiv \sum_\alpha v_\alpha^2\,, \qquad 
v^2=v_u^2+v_d^2 \simeq (246 \, {\rm GeV})^2 \,.
\eeq
Note that $\xi$
measures the misalignment of $v_\alpha$ and $\mu_\alpha$.

It is easy to check that $M^{(n)}$ possesses $2$ zero eigenvalues.
We shall identify the corresponding states with $2$
physical neutrinos of the Standard Model \cite{bgnn}, while
one neutrino acquires mass through mixing.
We can evaluate this mass by computing the product of the five
non-zero eigenvalues of $M^{(n)}$ [denoted below by
$\det' M^{\rm (n)}$]
\beq
\mbox{$\det'$} M^{\rm (n)} = 
m_Z^2 \mu^2 M_{\tilde \gamma}\cos^2\beta \sin^2\xi\,,
\eeq
where $M_{\tilde \gamma}\equiv \cos^2\theta_W M_1 + \sin^2\theta_W M_2$.
We compare this result with the product of the four neutralino masses
of the R-parity-conserving MSSM (obtained by computing the determinant
of the upper $4\times 4$ block of $M^{(n)}$ with $\mu_0$, $v_0$ 
replaced by $\mu$, $v_d$ respectively)
\beq
\det M^{\rm (n)}_0 = 
\mu\left(m_Z^2 M_{\tilde \gamma}\sin 2\beta-M_1 M_2 \mu\right).
\eeq
To first order in the neutrino mass, the neutralino masses are unchanged
by the R-parity violating terms, and we end up with \cite{enrico}
\beq \label{tree-level}
m_\nu = {\det' M^{\rm (n)} \over \det M^{\rm (n)}_0} =
{m_Z^2 \mu M_{\tilde \gamma}\cos^2\beta \sin^2\xi
\over m_Z^2 M_{\tilde \gamma}\sin 2\beta-M_1 M_2 \mu}\,.
\eeq
Thus, $m_\nu \sim m_Z \cos^2\beta \sin^2\xi$, assuming that all the relevant 
masses are at the electroweak scale.

A necessary and sufficient condition for $m_\nu\neq 0$ 
(at tree-level) is $\sin\xi\neq 0$, which implies that 
$\mu_\alpha$ and $v_\alpha$ are not aligned.
This is generic in RPV models.
In particular, the alignment of $\mu_\alpha$ and $v_\alpha$ is not
renormalization group invariant\cite{enrico,Hem}.
Thus, exact alignment at the low-energy scale 
can only be implemented by the fine-tuning of the model parameters.

\section{Sneutrino mass splitting}
In RPV low-energy supersymmetry, the sneutrinos mix with the Higgs
bosons.  Under the assumption of CP-conservation, we may separately
consider the CP-even and CP-odd scalar sectors.  
Consider first the case of one sneutrino generation. 
If R-parity is conserved, the CP-even scalar sector consists of two
Higgs scalars ($\hl$ and $\hh$, with $\mhl<\mhh$) and 
$\snu_+$, while the CP-odd scalar sector consists of 
the Higgs scalar, $\ha$, the Goldstone
boson (which is absorbed by the $Z$), 
and one sneutrino, $\snu_-$.  Moreover, the $\snu_\pm$ are mass
degenerate, so that the standard practice is to define eigenstates of
lepton number: $\snu\equiv (\snu_+ + i\snu_-)/\sqrt{2}$ and 
$\bar{\snu}\equiv\snu^*$.  When R-parity is violated, the sneutrinos
in each CP-sector mix with the corresponding Higgs scalars, and the
mass degeneracy of $\snu_+$ and $\snu_-$ is broken.
We expect the RPV-interactions to be small; thus, we can evaluate the 
resulting sneutrino mass splitting in perturbation theory.

The derivation of the CP-even and CP-odd scalar squared-mass matrices 
can be found in \Ref{GHRPV}. Working in the $v_m=0$ basis and for one 
generation we find
\beq \label{meven2sp}
M_{\rm even}^2= \pmatrix{
b_0\cot\beta+\quarter(g^2+g^{\prime 2})v_u^2 & 
-b_0-\quarter(g^2+g^{\prime 2}) v_u v_d & -b_1 \cr
-b_0-\quarter(g^2+g^{\prime 2}) v_u v_d &
b_0\tan\beta+\quarter(g^2+g^{\prime 2})v_d^2 & b_1\tan\beta \cr
-b_1 &  b_1\tan\beta & m_{\snu\snu^*}^2 \cr}\,,
\eeq
and
\beq \label{modd2sp}
M_{\rm odd}^2= \pmatrix{
b_0\cot\beta & b_0 & b_1 \cr
b_0 & b_0\tan\beta & b_1\tan\beta \cr
b_1 &  b_1\tan\beta & m_{\snu\snu^*}^2 \cr}\,,
\eeq
where 
\beq\label{moredefs}
m_{\snu\snu^*}^2\equiv 
(M^2_{\tilde L})_{11}+\mu_1^2-\eighth (g^2+g^{\prime 2})(v_u^2-v_d^2)\,.
\eeq
In the R-parity-conserving limit ($b_1=\mu_1=0$), one obtains the
usual MSSM tree-level masses for the Higgs bosons and the sneutrinos.

In both squared-mass matrices [\eqs{meven2sp}{modd2sp}], 
$b_1 \ll m_Z^2$ is a small parameter that can be treated
perturbatively.  We may then compute the sneutrino mass splitting
due to the small mixing with the Higgs bosons.  Using second order matrix 
perturbation theory to compute the eigenvalues, we find:
\beqa \label{snuonesnutwo}
&&\msnuone^2=m_{\snu\snu^*}^2+{b_1^2\over \cos^2\beta}\left[
{\sin^2(\beta-\alpha) \over (m_{\snu\snu^*}^2-\mhh^2)} + 
{\cos^2(\beta-\alpha) \over
(m_{\snu\snu^*}^2-\mhl^2)}\right]\,, \nonumber \\
&&\msnutwo^2=m_{\snu\snu^*}^2+
{b_1^2 \over (m_{\snu\snu^*}^2-\mha^2)\cos^2\beta}\,. 
\eeqa
Above, we employ the standard notation for the MSSM Higgs sector
observables \cite{hunter}.  Note that at leading order in
$b_1^2$, it suffices to use the values for the Higgs parameters in
the R-parity-conserving limit.
Then at leading
order in $b_1^2$ for the sneutrino squared-mass splitting,
$\dmsnutwo\equiv \msnuone^2-\msnutwo^2$ we find
\beq \label{dms}
\dmsnu = {2 \, b_1^2\, m_Z^2 \, m_{\snu\snu^*} \, \sin^2\beta \over
(m_{\snu\snu^*}^2-m_H^2) (m_{\snu\snu^*}^2-m_h^2) (m_{\snu\snu^*}^2-m_A^2)}\,.
\eeq
where $\dmsnutwo \simeq 2 m_{\snu\snu^*} \dmsnu$.

We now extend the above results to more than one generation of
sneutrinos.  In a basis where $v_m=0$,
the resulting CP-even and CP-odd squared mass matrices
are obtained from \eqs{meven2sp}{modd2sp} by replacing $b_1$ with
the three-dimensional vector $b_m$ and
$m^2_{\snu\snu^*}$ by the $3 \times 3$ matrix
\beq
\msnusnumn\equiv
({M^2_{\tilde L}})_{mn}+ \mu_m\mu_n-\eighth 
(g^2+g^{\prime 2})(v_u^2-v_d^2)\delta_{mn}\,.
\eeq
In general, $\msnusnumn$ is not expected to be flavor diagonal.  In this
case, the theory would predict sneutrino flavor mixing in addition to
the sneutrino--antisneutrino mixing exhibited above.  The relative
strength of these effects depends on the relative size of the RPV and
flavor-violating parameters of the model.  To analyze the resulting
sneutrino spectrum, we choose a basis in which squared-mass matrix
$\msnusnumn=(m^2_{\snu\snu^*})_m\delta_{mn}$ is diagonal.
In this basis $b_m$ is also suitably redefined.
(We will continue to use the same symbols for these quantities in the
new basis.)  The CP-even and CP-odd sneutrino mass 
eigenstates will be denoted by $(\snu_{+})_m$ and
$(\snu_{-})_m$ respectively.
It is a simple matter to extend the perturbative analysis
of the scalar squared-mass matrices if the 
$(m^2_{\snu\snu^*})_m$ are non-degenerate.  We then find that 
$(\dmsnutwo)_m \equiv  (\msnuone^2)_m- (\msnutwo^2)_m$ is given by
\eq{dms}, with the replacement of $b_1$ and $m^2_{\snu\snu^*}$ by 
$b_m$ and $(m^2_{\snu\snu^*})_m$, respectively:
\beq \label{dms-3}
(\dmsnu)_m = {2 \, b_m^2\, m_Z^2 \, (m_{\snu\snu^*})_m \, \sin^2\beta 
\over ((m_{\snu\snu^*}^2)_m-m_H^2) ((m_{\snu\snu^*}^2)_m-m_h^2) 
((m_{\snu\snu^*}^2)_m-m_A^2)}\,.
\eeq
Since \eq{dms-3} has been derived in the $v_m=0$
basis, it follows that in an arbitrary basis, all
sneutrino--antisneutrino pairs would be mass-degenerate if $b_\alpha$
and $v_\alpha$ were aligned.  However, this alignment is not
renormalization-group invariant.  Hence we expect that
all the sneutrino--antisneutrino pairs are generically
split in mass at tree-level.


\section{One-loop Neutrino masses}
In contrast to the sneutrino sector, only one 
neutrino mass is generated at tree-level due to neutrino mixing
with the neutralinos.
Masses for the remaining massless neutrinos
will be generated by one loop effects. There are two classes of one loop 
diagrams.
The first consists of fermion--sfermion loops
and depends on the RPV trilinear terms. The second, 
which in many cases is the dominant one, consists of 
sneutrino--neutralino loops and depends on the 
sneutrino--antisneutrino mass splitting. 
We now discuss both of these effects in turn.

First, consider the fermion--sfermion loops.
Contributions to the neutrino mass matrix are generated 
from diagrams involving a charged lepton-slepton loop
and an analogous down-type
quark-squark loop \cite{all}.
In the limit where the fermion masses can be neglected,
\beq \label{oneloopnu}
(m_\nu)_{qm} =
{1 \over 32\pi^2}
\left[
\sum_{\ell,p}\lambda_{q\ell p} \lambda_{mp\ell} m_\ell \sin 2\phi_\ell
\ln\left({M_{p_1}^2 \over M_{p_2}^2}\right) + 
3 \sum_{d,r} \lambda'_{qdr} \lambda'_{mrd} m_d \sin 2\phi_d
\ln\left({M_{r_1}^2 \over M_{r_2}^2}\right)
\right]
\,,
\eeq
where $\phi_{\ell}$ $(\phi_{d})$ is the mixing angle of the charge slepton 
(down type squark) squared-mass matrix, 
\beq \label{sinphi}
\sin2\phi_\ell = {2 A_\ell m_\ell \over 
\sqrt {(L^2-R^2)^2+4 A_\ell^2 m_\ell^2}}\,.
\eeq
Here, $A_\ell\equiv (A_E)_{0\ell\ell}-\mu_0\tan\beta$,
$L^2\equiv (M^2_{\widetilde L})_{\ell\ell} + 
(T_3 - e \sin^2\theta_W) m_Z^2 \cos 2 \beta$ and
$R^2\equiv (M^2_{\widetilde E})_{\ell\ell} + 
(e \sin^2\theta_W) m_Z^2 \cos 2 \beta$, 
with $T_3=-1/2$ and $e=-1$.
For $\sin2\phi_{d}$, take $e=-1/3$ and replace 
$M^2_{\widetilde E} \to M^2_{\widetilde D}$, 
$M^2_{\widetilde L} \to M^2_{\widetilde Q}$ and 
$\ell \to d$ in the above formulae.
%

Second, consider the sneutrino induced masses.
In general, the existence of a sneutrino--antisneutrino mass splitting, which
is a result of a $\Delta L=2$ interaction,
generates a one-loop contribution to the neutrino mass.
We have computed exactly the one-loop contribution to the neutrino mass
[$m_\nu^{(1)}$] from neutralino/sneutrino loops \cite{GH}.
In the limit of
$m_\nu,\Delta m_{\snu}\ll m_{\snu}$,
the formulae simplify, and we find in the one generation case
\beq \label{loopmass}
m_\nu^{(1)} = {g^2\Delta m_{\snu} \over 32 \pi^2 \cos^2 \theta_W}
\sum_j \,f(y_j) |Z_{jZ}|^2\,,
\eeq
where
$f(y_j) =  \sqrt{y_j}\left[y_j-1-\ln(y_j)\right]/(1-y_j)^2$, with
$y_j \equiv {m_{\snu}^2/m_{\tilde\chi^0_j}^2}$, and
$Z_{jZ}\equiv Z_{j2}\cos\theta_W-Z_{j1}\sin\theta_W$ is the
neutralino mixing
matrix element that projects out the $\widetilde Z$ eigenstate from the $j$th
neutralino. One can check that $f(y_j)<0.566$, and
for typical values of $y_j$ between 0.1 and 10, $f(y_j) > 0.25$.
Since $Z$ is a unitary
matrix, we expect as a rough order of magnitude estimate
\beq \label{impor}
m_\nu \sim  10^{-3} \Delta m_{\snu}\,. 
\eeq
In the three-generation model, a similar estimate holds for 
the loop contribution to each neutrino mass.

\section{The Neutrino spectrum}
The neutrino spectrum is determined by the relative size of the different
RPV couplings that control the three sources of neutrino masses. In the
$v_m=0$ basis these are $\mu_m$ [for the tree level mass, \eq{tree-level}], 
$b_m$ [for the
sneutrino induced one loop masses, \eqs{impor}{dms-3}] and $\lambda_{ijk}$ 
and $\lambda'_{ijk}$
[for the trilinear RPV induced one loop masses, \eq{oneloopnu}].
Therefore, in order to understand the structure of 
the neutrino spectrum we must have
a framework that predicts the magnitude of these parameters. Here
we give one example: models based on an abelian horizontal
symmetry \cite{LLL}. 
Further details and examples will be given in \Ref{ghprep}.

Consider a simple model based on a $U(1)$ abelian horizontal
symmetry, which is described in detail in \Ref{LLL}.
For our purposes, it is sufficient to mention that
the order of magnitude of all the model parameters
are determined by the assigned horizontal charges to the various
fields in the theory.  Moreover, many ratios are predicted
independently of the specific charge assignment.  For the parameters
relevant for neutrino masses we find the following ratios \beq {b_i
\over v \mu_i} \sim 1, \qquad {\lambda_{ijk} v \over \mu_i} \sim
{m^\ell_{jk} \over v}, \qquad {\lambda'_{ijk} v \over \mu_i} \sim
{m^d_{jk} \over v}, \eeq where $m^\ell_{jk}$ $(m^d_{jk})$ is the
charged lepton (down-type quark) mass matrix.  Inserting the above
order of magnitude predictions into eqs.~(\ref{tree-level}),
(\ref{impor}), (\ref{dms-3}) and (\ref{oneloopnu}) we find that the
tree level yields the dominant contribution to the heaviest neutrino
mass.  The two lighter neutrinos get their masses mainly from the
sneutrino--neutralino loops (as a consequence of the non-zero
tree-level sneutrino--antisneutrino mass splitting); 
these will be referred to as the sneutrino--induced masses.
The trilinear
RPV induced masses are suppressed compared to the sneutrino--induced
masses by at least a factor of order $30 \, m_b^4/v^4 \sim 10^{-6}$ and
are therefore completely negligible.  Moreover, the following
relations hold \beq
\label{ratio-3i} {m_i \over m_3} \sim 10^{-3} \sin^2\theta_{i3} , \qquad
{m_1 \over m_2} \sim \sin^2\theta_{12} .  \eeq

The fact that the sneutrino--induced neutrino mass
is the dominant radiative effect 
is not unique to the above example.
For example, in models with high
energy alignment, we also find that these contributions are larger than the
one-loop contribution from fermion--sfermion loops \cite{ghprep}.

\section{Summary and Conclusions}
Recent experimental
signals of neutrino masses and mixing may provide the first glimpse of
the lepton-number violating world.  In R-parity violating 
supersymmetric models that incorporate lepton number
violation, the sneutrino--antisneutrino mass spectrum may provide additional
insight to help us unravel the mystery of neutrino masses and mixing.
Such models generally predict a non-trivial neutrino spectrum in which 
there are several sources for neutrino masses. One
neutrino acquires mass at tree level via neutrino--neutralino
mixing. The other two neutrinos acquire radiative masses at one loop.
In many models, the dominant contribution of the radiative neutrino
masses is induced by the non-zero sneutrino--antisneutrino mass splitting.
A detailed study of sneutrino phenomenology at future colliders 
can complement the present day study of neutrino mass and mixing in
order to shed light on the nature of the underlying lepton-number violation.


\end{document}  